\documentclass[aps,prl,twocolumn]{revtex4}

\begin{document}

{\bf Comment on ``Magnetic response of Disordered Metallic Rings: 
     Large Contributions of Far Levels''}


In a recent Letter \cite{Schechter03}, Schechter {\em et al.}\ re-considered
the average magnetic response of
disordered metallic rings, on the basis of a
calculation first-order in the interaction (assumed to be phonon mediated,
hence attractive), for zero temperature, and for vanishing magnetic flux.
Their result is
\begin{equation}
  \chi (0) =  
  \frac{8 \pi \lambda E_{\rm Th}}{\Phi_0^2} {\cal M}^* , \;
  {\cal M}^* = \ln(\frac{E^*}{d}) ,
\end{equation}
where $\chi(\Phi) = dI(\Phi) /d\Phi$ is the susceptibility, $I(\Phi)$ 
the persistent current per ring, $\Phi$ the threading magnetic flux, 
$\lambda (<0)$ the dimensionless interaction constant,
$E_{\rm Th} = \hbar D/L^2$ the Thouless energy, $\Phi_0 =h/2e$ the flux
quantum, $d$ the mean level spacing, and $E^*$ a cut-off energy, given by
the minimum of $\hbar\omega_D$ and $\hbar /\tau$. In contrast, earlier
results \cite{Ambegaokar90} indicate that $E^* \sim E_{\rm Th}$ instead;
hence Eq.\ (1) suggests an ``increase'' of the susceptibility by 
a factor of about 4 (for typical experimental parameters).

In view of unresolved questions in relation to persistent currents,
see {\em e.g.} 
\cite{Levy90,Jariwala01,Deblock02}, concerning
the sign, the magnitude, and the temperature dependence,
we agree that further studies of the interaction
contribution are important. We doubt, however, that a first-order
calculation based on a reduced Hamiltonian 
can give reliable answers.
At least this approach must be contrasted with
standard many-body calculations \cite{Scalapino69,Altshuler85} which
support the approach in \cite{Ambegaokar90}, as detailed in 
\cite{Eckern91,Ambegaokar91}.

Let us recall the expression derived in \cite{Ambegaokar90,Remark} 
(see also \cite{Schmid91})
for the grand potential,
\begin{equation}
  \Omega (\Phi) = 2\lambda^* \sum_q T\sum_{\omega >0} 
  \frac{\omega}{\omega + D q_\Phi^2} ,
\end{equation}
where $q =2\pi n/L$, $n = 0, \pm 1, \dots$, 
$q_\Phi = q + (2\pi /L)\Phi /\Phi_0$, and
$\omega$ the Matsubara (Bose) frequencies. The applicability of
Eq.\ (2) is subject to restrictions, implicit in its
derivation, namely $d \ll \omega, Dq^2 \ll 1/\tau$, which implies {\em e.g.}\
that the temperature $T$ must be larger than $d$ (and larger than the
superconducting $T_c$ for the attractive case). 
The coupling constant
$\lambda^*$ contains Hartree and Fock contributions, averaged
over the Fermi surface. The validity of the arguments leading to
Eq.\ (2) persists when the screened Coulomb interaction is replaced
by the phonon Green's function. In the latter case, an additional
cut-off for the frequency summation is provided by the Debye
frequency $\omega_D$. A careful analysis of the higher order terms
has been given in \cite{Eckern91} and \cite{Ambegaokar91}, for the
repulsive and the superconducting case, respectively. 
From Eq.\ (2),
the $m$-summation in the expansion of the persistent current,
$I(\Phi) = \sum_{m=1}^\infty I_m \sin (2\pi m \Phi /\Phi_0 )$,
where $I_m = I_1 /m^2$ for 
$T \ll E_{\rm Th}$, is cut off at $m^* \sim (E_{\rm Th}/T)^{1/2}$,
and we recover Eq.\ (1), however, with 
${\cal M}^* =  \ln(E_{\rm Th}/T)$. For the linear response to be
valid, this also implies $\Phi /\Phi_0 \ll 1/m^*$. 

On the other hand,
taking the $q=0$ term into account only \cite{Schechter03}, and expanding
Eq.\ (2) for small flux, we obtain ${\cal M}^* =  \ln(E^* /T)$ with
$E^*$ the appropriate cut-off for the frequency sum, 
$\sim 1/\tau$ or $\sim\omega_D$. The 
connection with the results of \cite{Schechter03} is apparent when
$T$ approaches $d$. Clearly, with such a 
procedure, the flux periodicity is lost.

We emphasize that the wave-vector $q$, appearing above, is
the {\em sum} of the incoming momenta, but the 
interaction depends on the momentum {\em transfer} \cite{Ambegaokar90};
hence the relevant scale is set by $p_F$, and 
not by $\omega_D /v_F$, as argued in \cite{Schechter03}; see
\cite{Scalapino69}. Thus there is
no convincing argument which could justify singling out the $q=0$
contribution in Eq.\ (2), and we conclude that the reduced
BCS Hamiltonian leads to erroneous results in the present case.
(Taking $q=0$ only is valid
in a superconductor above but close to $T_c$, such that
the coherence length is larger than the system size \cite{Ambegaokar91}
-- but then terms of infinite order have to be summed.)
Nevertheless one can imagine starting with an effective
Hamiltonian, in which the interaction $V ({\bf p}-{\bf p}^\prime )$
is replaced by some ${\tilde V}({\bf p}-{\bf p}^\prime, q)$, and then
calculate the grand potential in first order in ${\tilde V}$. This
leads to Eq.\ (2) where, however,
$\lambda^*$ depends on $q_\Phi$, thereby 
guaranteeing the flux-periodicity of the results. Clearly, upon 
differentiation, ambiguous results are obtained, depending on the
choice of ${\tilde V}$. On the other hand, the replacement
$\lambda^* (q_\Phi) \to \lambda^* (q)$ at some arbitrary point in the
calculation has no foundation either.

Support from NSF (DMR-0242120), 
DAAD, and DFG (SFB 484) is acknowledged.

\medskip

U.\ Eckern and P.\ Schwab \\
Institut f\"ur Physik, Universit\"at Augsburg\\
86135 Augsburg, Germany

V.\ Ambegaokar \\
Laboratory of Atomic and Solid State Physics\\
Cornell University, Ithaca, New York 14850, USA

\medskip





\begin{references}
\bibitem{Remark}
  We use $\lambda^*$ here, which corresponds to
  $N(0)\overline{V}$ of \cite{Ambegaokar90}. In the following,
  $\hbar = k_B = 1$.
\end{references}
\end{document}